\documentclass[conference]{IEEEtran}

\usepackage[table]{xcolor}
\usepackage[most]{tcolorbox}

\usepackage{mathptmx} 

\usepackage{fancyhdr}
\usepackage[normalem]{ulem}
\usepackage[hyphens]{url}
\usepackage[sort,nocompress]{cite}
\usepackage[final]{microtype}
\usepackage{flushend}
\usepackage[bookmarks=true,breaklinks=true,letterpaper=true,colorlinks,citecolor=blue,linkcolor=blue,urlcolor=blue]{hyperref}
\usepackage{pifont}
\usepackage{comment}
\usepackage{xspace}
\usepackage{hyperref}
\usepackage{makecell}
\usepackage{subcaption}
\usepackage{pifont}
\usepackage{multirow}
\usepackage{algorithm,algorithmic}
\usepackage{arydshln}
\usepackage{cite}
\usepackage{amsmath,amssymb,amsfonts}
\usepackage{graphicx}
\usepackage{textcomp}
\usepackage{xcolor,soul}
\usepackage[hyphens]{url}
\usepackage{tikz}
\usepackage{array}

\usepackage[export]{adjustbox}

\definecolor{myred}{RGB}{153,0,0}
\definecolor{myblue}{RGB}{0,76,153}
\definecolor{mygreen}{RGB}{0,102,51}
\definecolor{mypurple}{RGB}{76,0,153}

\newcommand{\squishlist}{
 \begin{list}{$\bullet$}
  { \setlength{\itemsep}{0em}
     \setlength{\parsep}{0em}
     \setlength{\topsep}{1em}
     \setlength{\partopsep}{0em}
     \setlength{\leftmargin}{0em}
     \setlength{\labelwidth}{0}
     \setlength{\labelsep}{1} } }

\newcommand{\squishend}{
  \end{list}  }

\newcommand{\alphacrypt}{AlphaEvolve}
\newcommand{\ours}{\alphacrypt\xspace}

\hypersetup{final}

\renewcommand{\vec}[1]{\mathbf{#1}}

\newif\ifcommenton
\commentonfalse

\ifcommenton
\newcommand{\TODO}[1]{\textcolor{red}{[TODO] #1}}
\newcommand{\YC}[1]{{\color{orange}\bfseries [Yangyu: #1]}}
\newcommand{\AB}[1]{{\color{olive}\bfseries [Abhi: #1]}}
\newcommand{\alexey}[1]{\textcolor{navy}{[Alexey: #1]}}
\newcommand{\alind}[1]{\textcolor{codegreen}{[Alind: #1]}}
\newcommand{\TK}[1]{{\color{violet}\bfseries [Tushar: #1]}}
\newcommand{\DP}[1]{{\color{green}\bfseries [David: #1]}}
\newcommand{\HEO}[1]{{\color{blue}\bfseries [Taekyung: #1]}}

\newcommand{\fixme}[1]{{{\color{blue} #1}}}

\newcommand{\aferr}[1]{\textcolor{purple}{[Andrew: #1]}}
\newcommand{\afnote}[1]{{\color{teal}\textbf{aferr@:} \emph{#1}}}
\newcommand{\baiyu}[1]{\textcolor{teal}{[Baiyu: #1]}}
\newcommand{\jknote}[1]{{\color{blue}\textbf{Jon's note:} \emph{#1}}}
\newcommand{\jjnote}[1]{\jknote{#1}}
\newcommand{\shnote}[1]{{\color{magenta}\textbf{Shruthi:} \emph{#1}}}
\newcommand{\JT}[1]{{\color{brown}\bfseries [Jianming: #1]}}
\else
\newcommand{\TODO}[1]{}
\newcommand{\aferr}[1]{}
\newcommand{\afnote}[1]{}
\newcommand{\baiyu}[1]{}
\newcommand{\jjnote}[1]{}
\newcommand{\shnote}[1]{}
\newcommand{\JT}[1]{}

\newcommand{\YC}[1]{}
\newcommand{\AB}[1]{}
\newcommand{\alexey}[1]{}
\newcommand{\alind}[1]{}
\newcommand{\TK}[1]{}
\newcommand{\DP}[1]{}
\newcommand{\HEO}[1]{}
\newcommand{\fixme}
\fi

\newcommand{\tfhebootstrap}[1]{\texttt{TFHE-bootstrap}}
\newcommand{\hemul}[1]{\texttt{CKKS-Mult}}
\newcommand{\herot}[1]{\texttt{CKKS-Rot}}

\newcommand{\secref}[1]{\S\ref{#1}}
\newcommand{\figref}[1]{Fig.~\ref{#1}}

\def\BibTeX{{\rm B\kern-.05em{\sc i\kern-.025em b}\kern-.08em
    T\kern-.1667em\lower.7ex\hbox{E}\kern-.125emX}}

 \author{
      \IEEEauthorblockN{Shruthi Gorantala, Jianming Tong$^{\dagger}$, Asra Ali, Baiyu Li, Jonathan Katz, Jeremy Kun \\
         Thomas Steinke$^{\ddagger}$, Abhradeep Thakurta$^{\ddagger}$, Julian Walker$^{\ddagger}$, Amir Yazdanbakhsh$^{\ddagger}$}
      \IEEEauthorblockA{Google  \qquad  $^\dagger$Georgia Institute of Technology \qquad $^\ddagger$Google DeepMind}
  }
  \title{Adapting AlphaEvolve to Optimize Fully Homomorphic Encryption on TPUs}

\IEEEoverridecommandlockouts
\begin{document}
\maketitle
\thispagestyle{plain}
\pagestyle{plain}

\begin{abstract}
The deployment of Fully Homomorphic Encryption (FHE) at scale is hindered due to its heavy computational overhead.
While specialized hardware accelerators like Google Tensor Processing Units (TPUs) can help, mapping complex cryptographic kernels onto such architectures remains a challenge. 
%
%
Efficient execution requires  co-optimization between the systolic array-based Matrix Multiplication Unit (MXU) and Vector Processing Units (VPUs), as well as the orchestration of data movement across the vector register files. 
%
%
Existing compiler stacks often abstract low-level hardware utilization, requiring developers to adopt a manual trial-and-error process that often results in fragmented execution and underutilized resources.
%

To accelerate this development process, we use AlphaEvolve to automate the exploration of hardware-aware cryptographic-kernel optimizations.
We frame optimization as an evolutionary search problem, utilizing the closed-loop system provided by AlphaEvolve, that leverages LLM-driven code generation. We  use real-world feedback from hardware execution and rigorous correctness testing to guide the evolution process.
%
We evaluate \ours optimization on primitives for both the TFHE (Jaxite) and CKKS (CROSS) FHE schemes on Google Cloud TPUv5e, a contemporary TPU architecture.
Within 24 hours of automated exploration, \ours discovered implementation-level optimizations 
that improve TFHE bootstrap latency by 2.5$\times$ and CKKS rotation and multiplication latency by 1.31$\times$ and 1.18$\times$, respectively, 
relative to human-engineered state of the art.
%
These results demonstrate that \ours can be used to enable researchers to navigate the optimization trade-offs between cryptography, compilers, and hardware accelerators.
%
%
\end{abstract}
\section{Introduction}
\label{sec:intro}
The rapid proliferation of personal data in AI ecosystems has created a demand for privacy-enhancing technologies (PETs). 
Fully Homomorphic Encryption (FHE)~\cite{fhe_craig} stands as a foundational Privacy Enhancing Technology (PET) by enabling arbitrary computations on encrypted data without requiring decryption. 
Despite its privacy guarantees, Fully Homomorphic Encryption (FHE) incurs orders-of-magnitude compute and memory overheads, severely hindering its scalable deployment \cite{fhe_cacm}. To mitigate these bottlenecks, the community has increasingly pivoted toward specialized accelerators like Tensor Processing Units (TPUs). 


Mapping FHE primitives to accelerators requires engineering optimizations that spans across tiling workloads to strictly align with the underlying vector register granularity, navigating frontend abstractions (e.g., JAX and Pallas APIs for TPUs) to bypass costly, implicit layout transformations and sizing tensor granularities to effectively reduce off-chip memory latency.
Modern compiler stacks like XLA~\cite{openxla_github} abstract these microarchitectural details. However performance tuning then becomes a slow trial-and-error tuning loop where high-level code modification must traverse the compilation pipeline to observe its hardware impact. 
Current systems lack a systematic approach to map transformed cryptographic primitives onto commodity hardware designed for low-precision, GEMM-heavy workloads.
%
%
%

AI coding agents such as AlphaEvolve\cite{novikov2025alphaevolve} have demonstrated significant advances in multiple fields including algorithm design, improving code efficiency, architecture discovery, and enabling discovery across scientific domains. Further, the use of AI to improve cryptographic implementation remains largely unexplored. This is primarily because LLM hallucinations might introduce security vulnerabilities which are unacceptable in cryptographic implementations.

In this paper, we use \ours, an agentic framework, to optimize FHE kernels on hardware accelerators, such as Google's TPU, with rigorous correctness checking. 
We formulate kernel optimization as an evolutionary search problem using AlphaEvolve, with real-world hardware feedback. 
By automating the discovery of micro-architectural improvements, \ours enables cryptographic developers to explore architectural ideas and co-optimize cryptographic primitives on specialized hardware. 
The main contributions of our work are: 
\begin{itemize}
    
\item We formulate JAX/Pallas-level FHE-on-TPU kernel optimizations as an evolutionary search problem.  

\item  We design a hardware-in-the-loop evaluator that scores candidates using real TPU latency while rejecting candidates that fail robust functional correctness tests.

\item We identify concrete optimization patterns discovered by the search that outperform domain-expert baselines: loop unrolling to reuse common parameters, finer-grained scheduling to hide off-chip memory access latency, minimize datatype conversion, and XLA-favorable tiling that improves VReg utilization.

\item Within 24 hours of exploration, \ours achieved a 2.5$\times$ reduction in TFHE bootstrapping latency and speedups of 1.31$\times$ and 1.18$\times$ for CKKS rotation and multiplication, respectively.

\end{itemize}
\section{Background and Motivation}
\label{sec:bg}

\subsection{AlphaEvolve}
AlphaEvolve~\cite{novikov2025alphaevolve} is an agentic framework designed for algorithmic and scientific discovery by combining LLMs with the evolutionary search. It could be used to automatically evolve code in any programming language through a flywheel of code generation and code evaluation. 

The workflow begins with a human programmer configuring the system based on expert-guided, task specific instructions, an initial code example to seed the evolutionary database and an evaluator that scores the generated code. 

Next, the system samples code database and a pool of prompts. These prompts are fed into LLMs to generate new code modification. The generated code blocks are then compiled and fed into the evaluators, where they are scored, ranked, and only top-rank correct programs are added back to the evolutionary database. The ranking and evolution employs structured evolutionary techniques, such as island-based models or MAP-Elites algorithms~\cite{novikov2025alphaevolve}. These techniques balance genetic diversity based on inherent metrics (e.g., program length, code complexity) and performance on the evaluator metrics (fitness score). This optimization loop improves the code to find better program based on the fitness score.

\subsection{Fully Homomorphic Encryption}

FHE \cite{fhe_craig} enables computation on encrypted data. A homomorphic operation transforms ciphertexts so that the decrypted results are the same as if the operation were performed on the underlying plaintext data.
Existing FHE schemes can be broadly categorized into scalar schemes and vector schemes based on the type of data they encrypt. Scalar schemes such as TFHE \cite{tfhe} encrypt bits or short integers, and supports general-purpose computation on encrypted data. Vector schemes such as CKKS \cite{ckks} pack a vector of real or complex numbers into a ciphertext, and are widely used for privacy-preserving machine-learning and numerical workloads as vectors of data are processed in a SIMD fashion.
Both approaches lead to 2$\sim$5 order-of-magnitude memory and computation overhead.

In this paper, we use \alphacrypt{} to optimize implementations of both schemes on TPUs. 

\subsubsection{TFHE - \tfhebootstrap{}} This is the key primitive in TFHE used in Google transpiler \cite{transpiler} and HEIR compiler \cite{ali2025heir}. TFHE bootstrap algorithms utilize the so-called RingGSW scheme defined over a polynomial residual ring $R_Q = \mathbb{Z}_Q[X]/(X^{n}+1)$ and a special, cryptographic accumulator to perform decryption homomorphically. The primary bottleneck in the bootstrap operation is \texttt{blind\_rotate}, which consists of a sequence of multiplication-accumulation operations on polynomials. Note that we reformulate polynomial multiplications in $R_Q$ as Toeplitz matrix-vector products to take advantage of MXUs in TPUs.

\begin{figure}[h]
  \centering
  \footnotesize 
  \begin{tabular}{p{0.98\columnwidth}}
    \hline
    \textbf{Algorithm:} $\mathsf{ToeplitzMul}(\vec{a}, \vec{b})$ \\
    \textbf{Input:} $\vec{a} = (a_0, \dots, a_{n-1})$, $\vec{b} = (b_0, \dots, b_{n-1})$ \\
    \hline
    
    Construct a Toeplitz matrix $T(\vec{a}) \in \mathbb{Z}^{n \times n}$: \\
    $\quad T(\vec{a}) = \begin{bmatrix} 
    a_0 & a_1 & \cdots & a_{n-1} \\ 
    -a_{n-1} & a_0 & \cdots & a_{n-2} \\ 
    \vdots & \vdots & \ddots & \vdots \\ 
    -a_1 & -a_2 & \cdots & a_0 
    \end{bmatrix}$ \\
    \quad \hfill \textit{\scriptsize // Note: Negacyclic entries account for reduction modulo $X^n + 1$}. \\

    Compute the matrix-vector product $\mathbf{c} = T(\vec{a}) \cdot \vec{b} \in \mathbb{Z}_q^n$. \\
    
    
    \textbf{return} $\vec{c}$ \hfill \textit{\scriptsize // Represents the polynomial $c(X) = \sum_{i=0}^{n-1} c_i X^i$.} \\
    \hline
  \end{tabular}
  \caption{Polynomial multiplication $a(X) \cdot b(X) \in R_Q = \mathbb{Z}_Q[X]/(X^n+1)$ via Toeplitz matrix-vector product, where $\vec{a}$ and $\vec{b}$ are coefficient vectors of $a(X)$ and $b(X)$, respectively.}
  \label{fig:toeplitz-multiplication}
\end{figure}


\subsubsection{CKKS - \herot{} and \hemul{}} 
They perform homomorphic slot rotation and slot-wise multiplication, respectively, which are the backbone components in almost all CKKS use cases.
We optimize their implementations from CROSS~\cite{tong2025CROSS}, combining automorphisms, key switching, rescaling, tensor multiplication, and modular reduction operations.

\subsection{TPU Execution Model}

\begin{figure}[t!]
    \centering
    \includegraphics[width=\columnwidth]{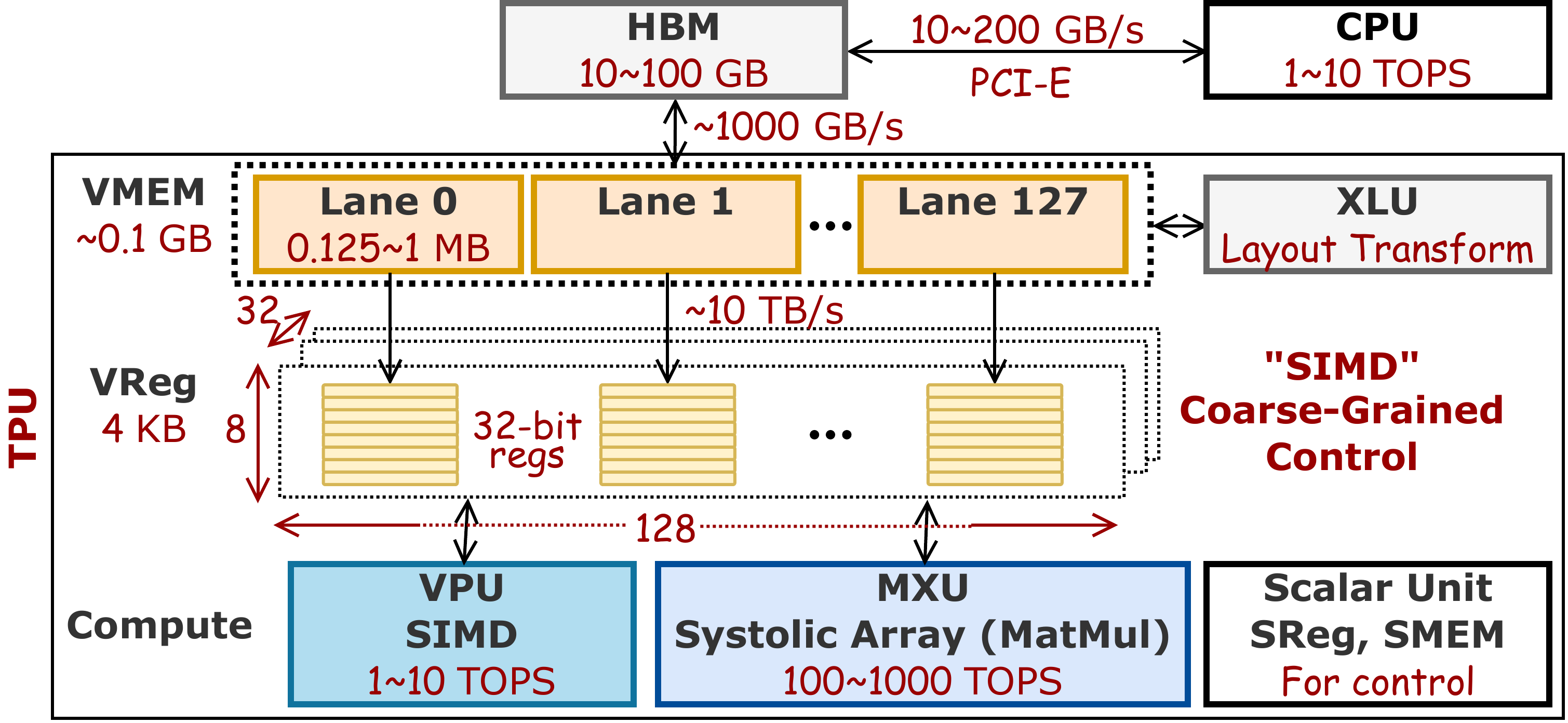}
    \caption{TPU microarchitecture overview -- SIMD machine.}
    \label{fig:tpu_micro_arch}
    \vspace{-5mm}
\end{figure}

\figref{fig:tpu_micro_arch} summarizes the TPU features. 
On-chip vector memory (VMEM) is partitioned across 128 lanes. Each lane contains 8 sublanes, and each sublane provides 32 32-bit registers. Registers are organized into $(8,128)$ blocks, called vector registers (VRegs), each formed by selecting one register from every sublane across all 128 lanes. The XLU performs data reorganization, while the VPU and MXU accelerate vector and matrix operations. Both computation and data movement execute in a single-instruction, multiple-data (SIMD) manner over one $(8,128)$ VReg at a time. Consequently, programs that fully utilize VRegs achieve higher efficiency, whereas operations requiring fine-grained or irregular data access incur significant overhead.

For FHE, the implication is direct: mathematically equivalent implementations can differ sharply in performance depending on shapes, layout and scheduling choices of JAX/Pallas tensors. 

\subsection{The TPU Compiler Stack}
We inherit the multi-layer TPU compilation stacks and profiling system to support FHE computations: 

\subsubsection{JAX} expresses high-level numerical and cryptographic computation. Both expert-crafted TPU libraries, Jaxite~\cite{jaxite_github} and CROSS~\cite{tong2025CROSS}, use JAX / Pallas as the frontend.

\subsubsection{Pallas} serves as a kernel language that allows for the creation of custom, low-level hardware kernels within JAX. Our system relies on Pallas kernels to establish fine-grained control over data movement and memory staging on the TPU~\cite{pallas_jax}.

\subsubsection{XLA} lowers JAX/Pallas programs to TPU executables and applies optimizations such as fusion and layout selection~\cite{he2023accelerated,openxla_github}. XLA compilation passes are shape-sensitive.

\subsubsection{Xprof} is an integrated profiling tool within the XLA ecosystem that captures high-fidelity hardware execution traces. It extracts metrics such as compute and memory bandwidth utilization as well as on-chip and off-chip storage occupancy.

\subsection{Challenge: Compiler-Obscured Performance Engineering}
TPU performance engineering demands rigorous optimizations: (1) tiling workloads to strictly align with VReg granularities, (2) navigating frontend abstractions to bypass implicit layout transformations, and (3) sizing tensors to effectively hide off-chip memory latency behind compute.

Developers are strictly bound to high-level programming interfaces such as JAX and Pallas, limiting the easy access to manipulating the low level schedule. At this level of abstraction, every source modification must traverse hundreds of opaque compilation passes within stacks like XLA~\cite{openxla_github} before finally materializing as a low-level TPU execution trace. Because these intermediate layers obscure critical microarchitectural details, adapting an algorithm for TPU execution devolves into a sluggish trial-and-error process. Developers are forced to experimentally deduce the specific shapes, casts, loops, and tiling parameters at frontend JAX/Pallas that cause XLA to produce a good TPU program.

This paper adapts \alphacrypt{} to automate this compiler-steering loop while preserving correctness and security guarantees.

The use of AI to improve cryptographic implementation is challenging as LLM hallucinations may introduce security vulnerabilities prone to side channel attacks. We use optimizing FHE as an initial case study where the data is encrypted. 

\section{Adapting AlphaEvolve for Cryptography}
\label{sec:design}

\begin{figure}[t!]
    \centering
    \includegraphics[width=0.90\linewidth]{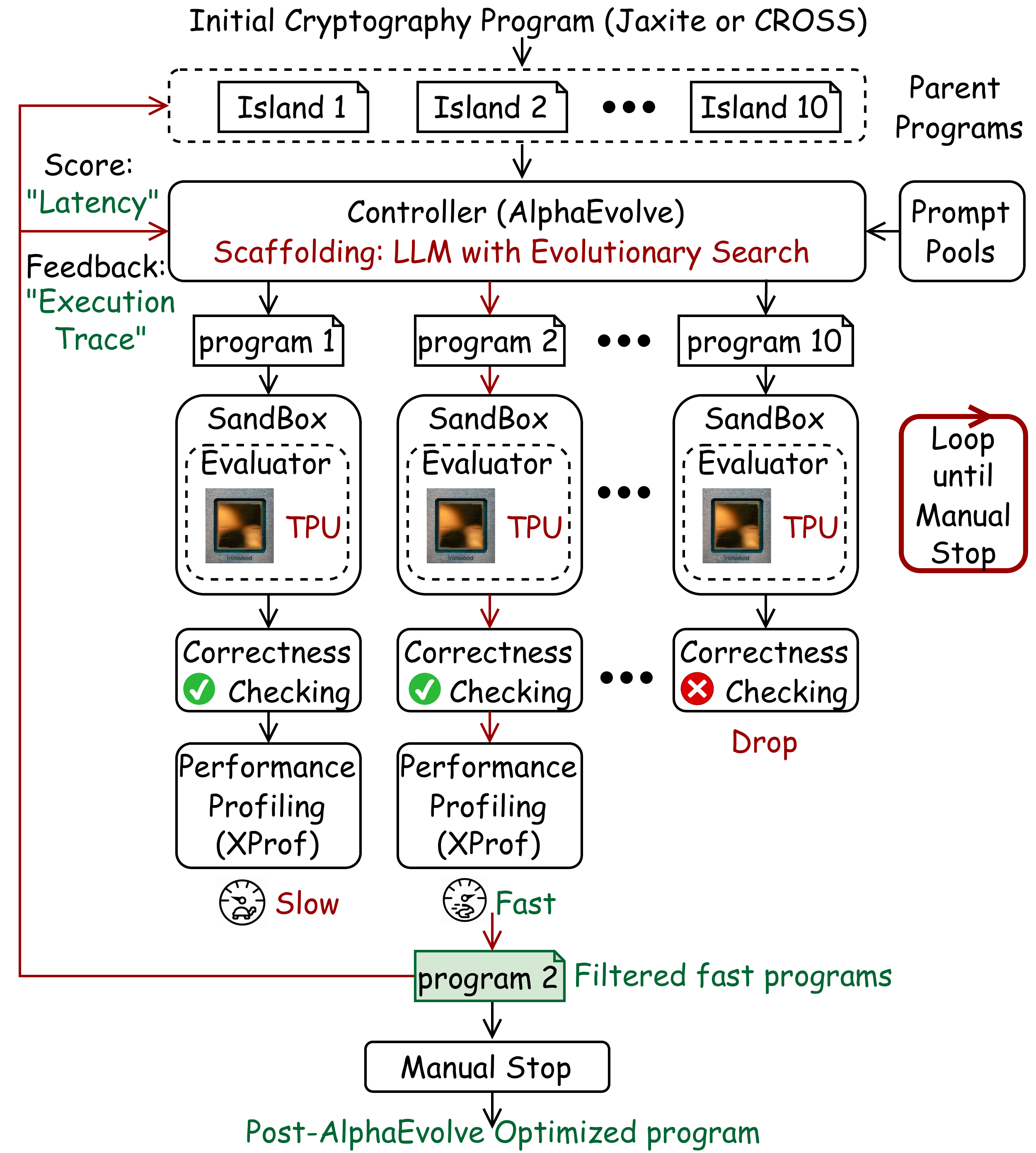}
    \caption{Workflow for adapting AlphaEvolve to cryptographic kernel optimization. AlphaEvolve samples parent programs, prompts an LLM to generate candidate JAX/Pallas implementations, and evaluates each candidate in a sandbox on TPU hardware. Invalid candidates are rejected by correctness and security checks. Valid candidates are scored by latency and profiled through execution traces. Fast candidates are returned to the evolutionary population for further search.}
    \label{fig:system_design}
\end{figure}

Configuring AlphaEvolve requires users to specify initial prompts (which establish a persona, algorithm and hardware platform), initial programs (which will be optimized by AlphaEvolve), the AlphaEvolve framework itself  and an evaluation strategy (which validates the correctness, and then feeds reward as feedback back to AlphaEvolve).

\figref{fig:system_design} provides an overview of various parts of the system with details described below:

\subsection{Initial Prompts}
The initial prompts describe the task and role of \alphacrypt{} as an expert in hardware-aware kernels engineering. 
The prompts also include a functional description of the cryptographic algorithms being optimized, architecture of the hardware accelerator, optimization strategies and a pool of potential optimization strategies.

\subsection{Initial Program}
We feed well-optimized reference codes into the evolutionary search framework as the initial programs. 
The choice of initial program needs a balance between providing enough code for \alphacrypt{} to optimize while not allowing it to change the security guarantees provided by the algorithm. 
We use co-evolution as a process to evolve multiple kernels in tandem to find kernel improvements within various aspects of the cryptosystem. 
In addition to co-evolution, co-optimization can be used to evolve security parameters and kernels, which allows for tuning security parameters for accelerator-specific kernel super-optimization.

\subsection{Evaluation Engine}
The evaluation engine only rewards candidate kernels that satisfy these multi-tiered correctness and security checks, thereby pruning invalid or insecure code early in the evolutionary search process.
\subsubsection{Correctness Checks}
To mitigate the risks of reward hacking and code hallucination inherent in LLM-generated kernels, we enforce strict correctness tests using randomized input vectors across all computational sub-layers. 
The correctness checking requires correct compilation to mitigate the risks of unintended bugs and erroneous code. 
This is done by testing subset of sub-modules as well as incorporating end-to-end tests which initializes the cryptosystem with valid security parameters, generates randomized inputs, encrypts the inputs, optionally runs cryptographic computations on the ciphertexts such as homomorphic computations for FHE cryptosystem, decrypts the resultant ciphertexts and checks against the expected plaintext outputs.

\subsubsection{Security Checks}
Our system validates any modifications to the parameters of the cryptosystem maintain standardized security guarantees (example: at least 128-bit security for Fully Homomorphic Encryption utilizing acceptable parameters generated from industry standard lattice estimators \cite{lattice_estimator}). 
Furthermore in case of fully homomorphic encryption, we implement safeguards to prevent the selection of scheme parameters that could lead to decryption failures by adding additional end-to-end test with high depth computation. 
\subsubsection{Target Hardware Latency as Reward} 
By deploying directly on the target platform, we use real-world latency as the primary reward score. 
In this paper, candidate kernels in Jaxite and CROSS are lowered through the XLA compiler and executed on physical TPU hardware.
Our reward score uses microsecond-level execution latency of the resulting cryptographic primitive, excluding XLA compilation overhead to isolate runtime performance. 

A persistent bottleneck in FHE acceleration is the lack of parity between isolated primitive speedups and actual end-to-end performance due to the bottlenecks introduced by data movement between host and device and also between devices. 
To ensure primitive kernel speedup translates to end-to-end speedup, \alphacrypt{} evolves the one or more primitive kernels simultaneously, while using end-to-end latency across the entire functional module as the score. 
Measuring end-to-end latency proves to be a useful reward metric for co-evolution and co-optimization techniques.

\subsubsection{Feedback}
To facilitate the autonomous discovery of hardware-optimal execution patterns, our system captures high-fidelity execution traces during the evaluation cycle using xprof integrated with XLA. 
These metrics from profiling including compute and memory bandwidth utilization as well as on-chip and off-chip storage occupancy are fed back to AlphaEvolve as diagnostic feedback. 
By analyzing these traces, the agent can algorithmically identify latency bottlenecks and instances where the XLA compiler needs additional signal through JAX (or Pallas), enabling targeted iterative refinements to the cryptographic kernels.

\subsection{Manual reviews / Functional equivalence}
The best program generated by the evolutionary loop can be integrated into the original library after human review. Currently, a human programmer reviews the best program and merges the improved code into the original library. To reduce human-in-the-loop reviews, in principle a module to check functional equivalence of the initial code and the AlphaEvolve generated code could be used, though doing this is currently an unsolved research problem.

\subsection{Putting it all together}
After configuring initial prompts, initial programs, evaluation engine, and setting up appropriate hardware environment for the evaluators, \alphacrypt{} begins the discovery process of automatically optimizing the cryptographic kernels. 
After reaching convergence based on the fitness function, the best generated program can be further tested for correctness through human reviews.
The final code can then be merged into the original library and deployed for use. 

\section{AlphaEvolve for optimizing FHE}
\label{sec:ckks_results}

\subsection{Setup}

\subsubsection{Initial programs}
For TFHE, we use the bootstrap implementation from jaxite \cite{jaxite_github} and allow AlphaEvolve to modify various parts of \texttt{blind\_rotate}. 
For CKKS, we adopt  \herot{} and \hemul{} from the TPU-based CKKS library, CROSS~\cite{tong2025CROSS} that offers state-of-the-art throughput, as the initial programs and apply AlphaEvolve to search for faster implementations. 
We expose the implementations of \textsc{Rescale}, \textsc{TensorMultiply}, \textsc{KeySwitch}, \textsc{ApproxModDown}, and \textsc{AutoMorphism} from both kernels to AlphaEvolve.

\subsubsection{Prompt Setup}
We use the system prompt together with a prompt sampled uniformly from \secref{sec:appendix_prompts}. These prompts are designed to steer AlphaEvolve towards implementation-level improvements, especially scheduling and layout choices that increase VReg utilization in TPU and minimize implicit layout transformation, rather than toward new algorithms discovery.

\subsubsection{Correctness, and Performance Tests}
For TFHE, we incorporate end-to-end tests where the cryptosystem is initiated with parameters offering 128-bit security and random inputs are encrypted and evaluated against the generated code. The outputs are then decrypted and compared against the expected plaintext values. All values of the 3-bit lookup tables are evaluated for correctness.

For CKKS, functional correctness is enforced using a hidden test set with fixed ciphertext inputs and expected ciphertext outputs.

For both schemes, performance is measured from XProf traces, and we return the negative kernel latency as the optimization score so that lower latency corresponds to a higher score.

We run the evolution process with 10 controllers and 10 evaluators per controller, for a total of 100 TPUv5e chips.

\begin{table}[!htp]\centering
\vspace{-2mm}
\caption{Latency Comparison to SoTA\cite{tong2025CROSS,jaxite_github}}
\vspace{-1mm}
\label{tab:tpu_optimization_steps}
\resizebox{\columnwidth}{!}{
\begin{tabular}{c|c|c}
\hline
\textbf{Workload} & \textbf{Latency Improvement} & \textbf{Code/Analysis} \\
\hline
Blind Rotation (TFHE) & 9.4 ms (TPUv5e-1) & Code: jaxite~\cite{jaxite_github} \\
+ Unrolling & 10 $\rightarrow$ 7.8 ms & Code: \cite{jaxite_pr_88_unroll}  Reason: \secref{ref:unroll} \\
+ Fine-grained scheduling & 7.8 $\rightarrow$ 6.3 ms & Code: \cite{jaxite_pr_89_vector_polymul} Reason: \secref{ref:fine_grained}\\
+ Type Cast Removal & 6.3 $\rightarrow$ 3.5 ms & Code: \cite{jaxite_pr_90_single_bfloat16} Reason: \secref{ref:data_type} \\
\rowcolor[HTML]{E6EFDB}
\textbf{Post-AlphaEvolve blind Rotation} & \textbf{9.4 ms $\rightarrow$ 3.5 ms} & {\color{myred}{\textbf{2.85$\times$ speedup}}} \\
\rowcolor[HTML]{E6EFDB}
\textbf{Post-AlphaEvolve bootstrap} & \textbf{10 $\rightarrow$ 4 ms} & {\color{myred}{\textbf{2.5$\times$ speedup}}} \\
\hline \hline
\hemul{} (CKKS) & 6340 \textmu s (TPUv5e-1) & Code \href{https://github.com/EfficientPPML/CROSS/blob/a163ca741a3385a4102322220ab700440089affb/jaxite_word/hemul.py}{\scriptsize$\nearrow$} \cite{tong2025CROSS} \\
+ XLA-favor tiling choice & 5359 \textmu s (TPUv5e-1) & Code: \href{https://github.com/EfficientPPML/CROSS/blob/v2.0.0/jaxite_word/hemul.py}{\scriptsize$\nearrow$} Reason: \secref{ref:cross_reason} \\
\rowcolor[HTML]{E6EFDB}  \textbf{Post-AlphaEvolve \hemul} & \textbf{6340 $\rightarrow$ 5359 \textmu s} & {\color{myred}{\textbf{1.18$\times$ speedup}}} \\
\hline \hline
\herot{} (CKKS) & 4854 \textmu s (TPUv5e-1) & Code: \href{https://github.com/EfficientPPML/CROSS/blob/a163ca741a3385a4102322220ab700440089affb/jaxite_word/herot.py}{\scriptsize$\nearrow$} \cite{tong2025CROSS} \\
+ XLA-favor tiling choice & 3714 \textmu s (TPUv5e-1) & Code: \href{https://github.com/EfficientPPML/CROSS/blob/v2.0.0/jaxite_word/herot.py}{\scriptsize$\nearrow$} Reason: \secref{ref:cross_reason} \\
\rowcolor[HTML]{E6EFDB} \textbf{Post-AlphaEvolve \herot} & \textbf{4854 $\rightarrow$ 3714 \textmu s } & {\color{myred}{\textbf{1.3$\times$ speedup}}} \\ \hline
\end{tabular}}
\vspace{-5mm}
\end{table}

\subsection{AlphaEvolve for TFHE}

AlphaEvolve explored several distinct classes of optimizations, ranging from standard loop transformations to complex kernel scheduling and data type optimizations.
  
\subsubsection{Loop Unrolling for Parameter Reuse\cite{jaxite_pr_88_unroll}} \label{ref:unroll}
  In the negacyclic vector-matrix polynomial multiplication (\texttt{negacyclic\_vector\_matrix\_polymul}) routine~\cite{jaxite_github}, the system independently discovered that unrolling the primary \texttt{for} loop by a factor of 8 yielded significant latency reductions. In the baseline trace, the loop executes in serial, loading the common parameters repeatedly. By unrolling the loop, common parameters get reused across multiple loop iterations. Because the target parameter size (3.51\,MB) fit within the available on-chip memory, this transformation reduced latency from 10 ms to 7.8 ms. This highlights that AlphaEvolve could automatically identify and exploit loop unroll optimizations.
  
\subsubsection{Fine-Grained Memory and Compute Scheduling\cite{jaxite_pr_89_vector_polymul}}
\label{ref:fine_grained}
  The system optimized data access patterns by adjusting the tiling and scheduling choice of the kernel. Specifically, it tiled one big JAX tensor into two blocks with shape halved. This modification allowed the XLA compiler to schedule more fine-grained data loading from off-chip memory, and hiding it behind computation. This optimization reduced the overall bootstrapping latency from 7.8\,ms to 6.3\,ms.
  
\subsubsection{Elimination of Redundant Type Casts\cite{jaxite_pr_90_single_bfloat16}}\label{ref:data_type}
  The most significant single optimization involved the removal of an explicit bitcast operation. The baseline implementation explicitly cast a right-hand side (\texttt{rhs}) operand to 8-bit integers (\texttt{i8}). AlphaEvolve identified that the dynamic range of the input data (values from 0 to 255 due to RLWE decomposition in \cite{jaxite_pr_90_single_bfloat16}) was already sufficiently captured by the existing \texttt{bf16} precision. By removing the redundant and expensive bitcast operator, the system reduced the execution time from 6.29\,ms to 3.35\,ms. This optimization requires a deep understanding of both the mathematical dynamics of TFHE and the specific hardware execution costs of bitcast operations, making it difficult for human engineers to consistently identify.

\subsection{AlphaEvolve for CKKS}
\label{ref:cross_reason}

\begin{itemize}
\item \herot{}:
On TPUv5e, AlphaEvolve reduces the latency of \herot{} from 4874 \textmu s to 3754 \textmu s, yielding a $1.31\times$ speedup. This performance gain is primarily achieved by partitioning a tensor into two smaller sub-tensors. This strategic split triggers specialized XLA optimizations that significantly improve Vector Register (VReg) utilization during the \textsc{AutoMorphism} stage. Consequently, \textsc{AutoMorphism}'s contribution to the total end-to-end latency drops from 12\% to just 4\%. These results demonstrate AlphaEvolve's ability to expose and exploit non-intuitive, compiler-specific performance ``sweet spots" that are exceedingly difficult for human engineers to discover manually.

\item \hemul{}:
On TPUv5e, AlphaEvolve reduces latency of \hemul{} from 6340\,\textmu s to 5359\,\textmu s, corresponding to a 1.18$\times$ speedup. Such performance improvement comes from the scheduling optimization to reuse common parameters. The modest improvement indicates that \hemul{} is less constrained by underlying compiler inefficiencies.

\end{itemize}
These results show that AlphaEvolve can improve CKKS kernel performance on TPU by discovering implementation changes that increase effective VReg utilization and reduce inefficiencies in the execution schedule. 

\subsection{Takeaways}
The three workloads show that AlphaEvolve is most effective for cryptographic kernels when the bottleneck is caused by a mismatch between cryptographic data layout and the TPU compiler/hardware execution grain.
In TFHE, the search discovers both conventional locality improvements and a non-obvious cast elimination that depends on the ciphertext decomposition range.
In CKKS, the search finds tensor shapes that steer XLA toward higher VReg utilization.
Across all cases, the successful edits preserve the correctness and security while improving how the algorithm is presented to the compiler. By using AlphaEvolve to expose the performance headroom in the current TPU deployment flow, we also identify a key software bottleneck: existing programmable APIs force tensor reshapes and layout reorganizations that are not inherent to the hardware, introducing avoidable overhead.

\begin{figure}[t!]
    \centering
    \includegraphics[width=\linewidth]{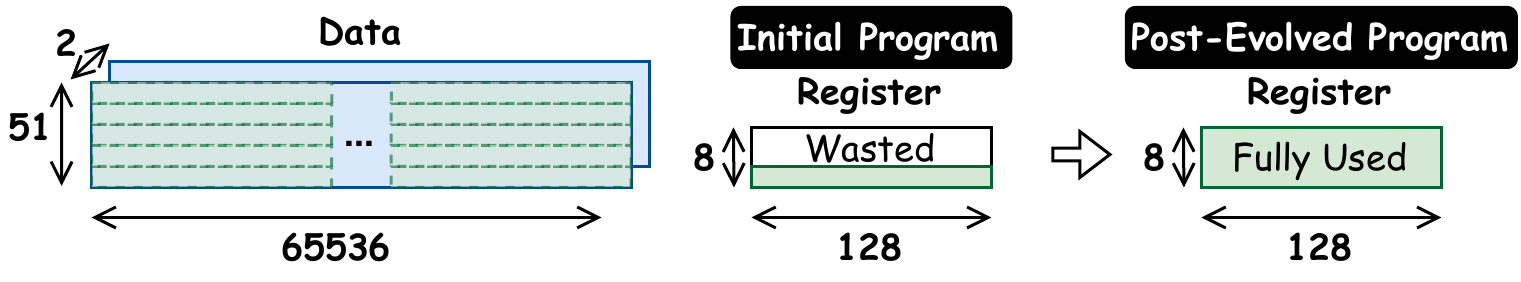}
    \vspace{-5mm}
    \caption{Effect of the discovered layout changes. The initial program leaves substantial VReg capacity unused because operations are expressed at a granularity that does not match VReg. The evolved program exposes fuller VReg utilization.}
    \label{fig:analysis}
    \vspace{-5mm}
\end{figure}

\section{Related Work}
\label{sec:comp}
To our knowledge this is the first application of AI technologies to improve cryptographic performance. The closest work for improving kernel generation is done by autocomp\cite{hong2025autocomp} and EvoX \cite{liu2026evoxmetaevolutionautomateddiscovery} where agentic infrastructure has been used to improve machine learning kernels. 
  \section{Conclusion}
  \label{sec:conclusion}
  This paper presents the first application of agentic AI to optimize FHE primitives on TPUs. 
  We detail the design of using \alphacrypt{}, a closed-loop system that combines LLMs with real-world hardware execution feedback and rigorous correctness checks.
  We demonstrate the framework's ability to autonomously discover performance improvements on TPUs. 

  Our results demonstrate that AI-driven search uncovers optimizations frequently missed by domain experts.
 
  %
  %
  
  \subsection{Experience and Lessons Learned using AlphaEvolve}
  \label{sec:experience}
   Some of our experiences and learnings in adapting AlphaEvolve for cryptographic optimization are described below:  
   
   \subsubsection{Granularity of Optimization} Targeting isolated operations, such as polynomial multiplication (\texttt{polymul} \cite{jaxite_github}), failed to produce end-to-end speedups. We found that the agent requires visibility into broader operational scopes to make a meaningful impact on overall latency.
   
   \subsubsection{Co-evolution and System-Level Scoring} Building on the need for broader visibility, shifting to the co-evolution of the \texttt{external\_product} and scoring it against the complete bootstrapping execution successfully reduced latency from 10\,ms to 5.6\,ms.
    
   \subsubsection{Guided Domain Constraints} While evolving the \texttt{blind-rotate} operation \cite{jaxite_github}, the system discovered that RLWE decomposition allowed input tensors to safely use \texttt{int8} rather than \texttt{int32}, and that unrolling the loop by a factor of eight yielded significant benefits. This underscores the importance of domain experts selectively granting the AI permission to modify high-impact code blocks.

   \subsubsection{Comprehensive Correctness Tests} Initial attempts to evolve the AND-gate failed to cover all values of a 3-input lookup table (\texttt{lut3} \cite{jaxite_github}). We had to update the evaluator to exhaustively verify all lookup table values, demonstrating that strict, comprehensive validation is essential to prevent the model from reward hacking.  

    \subsubsection{Security Checks} Because performance cannot come at the expense of cryptographic integrity, we introduced a standardized parameter check using a lattice estimator to ensure at least 128-bit security when co-optimizing scheme parameters. 
   
   \subsubsection{Feedback and Prompts} To effectively steer the evolutionary search, we integrated hardware profiling tools (\texttt{xprof}) directly into the feedback loop and provided hardware-specific optimization guidelines as automated compilation prompts.

  \subsection{Future Work}
  \label{sec:future_work}
  Our initial results focus on TPUs but the framework is  extensible and can potentially optimize FHE primitives as well as any cryptographic primitives on a variety of hardware accelerators. 
  It can also be integrated with simulators to guide specialized hardware-software co-design. 
  To ensure security guarantees are maintained during hardware optimization, future iterations can integrate parameter selection with noise tracking. 
  Furthermore, the bottleneck of manual code reviews can be reduced by adding functional equivalence checkers directly into the validation pipeline. 
  Finally, a fundamental open question remains in the design of reward functions to safely allow AI agents to discover novel cryptographic algorithms and protocols while preserving security guarantees.

\section*{Acknowledgments}
\label{sec:acknowledgments}
This work represents an equal contributions between the first two authors, with contributing co-authors listed alphabetically
The authors express their gratitude to Charles Hong, Andrew Ferraiuolo, Ben Kreuter, and Cindee Madison for their valuable feedback during various phases of this work. 
We thank Po-Sen Huang, Ng\^{a}n (NV) V\~{u}, and the AlphaEvolve team for their assistance in establishing the experimental framework. 
We thank Drew Angeloff, Ankita Malviya Bairaria, Naomi Black, Elie Burzstein, Bryant Gipson, Pankaj Rohatgi, Amanda Walker, Moti Yung and the Safeworks leadership team for their guidance throughout this project.
Finally, we also thank our extended team at Google DeepMind for their support of this research direction.

We also acknowledge the use of large language models (LLMs) in the preparation of this manuscript. 
Specifically, the models were used for proofreading, summarizing text, and generating LaTeX outlines. 
The authors have reviewed and verified all generated content and take full responsibility for the final manuscript.

\bibliographystyle{IEEEtranS}
\bibliography{reference}

\appendix


  
\subsection{LLM Instruction-Prompt Pool}
\label{sec:appendix_prompts}

\noindent $\bullet$ \textbf{Prompt 1:} Suggest a new way to minimize data reorganization in the program.

\noindent $\bullet$ \textbf{Prompt 2:} Suggest a new way to keep the logical shapes of JAX arrays invariant throughout execution.

\noindent $\bullet$ \textbf{Prompt 3:} Suggest a new way to hide off-chip memory latency behind computation.

\noindent $\bullet$ \textbf{Prompt 4:} Suggest a new optimization idea based on your expert knowledge of TPU performance tuning.

\subsection{Role and Hardware Architecture Prompts}

\noindent $\bullet$ \textbf{Role:} You are an expert hardware-aware compiler optimizer specializing in TPU architectures.
      
\noindent $\bullet$ \textbf{Objective:} Optimize the input computational graph for maximum performance on the TPU v5e architecture. You must make strategic decisions regarding operation mapping (Vector vs. Matrix units), memory hierarchy management, and parallelism.
      
\noindent $\bullet$ \textbf{Target Hardware Specifications:}
      \begin{itemize}
          \item \textbf{Compute Capabilities (Peak):}
          \begin{itemize}
              \item Clock Frequency: 1.50 GHz
              \item Matrix Units (MXU): 197 TFLOPS (\texttt{bf16}/\texttt{fp8}), 394 TOPS (\texttt{int8}), 788 TOPS (\texttt{int4}).
              \item Vector Units (VPU): \texttt{vmatmul} (1.54 TB/s), \texttt{vmatpush} (6.16 TB/s).
          \end{itemize}
          
        \item \textbf{Memory Constraints (CRITICAL):}
          \begin{itemize}
              \item \texttt{CMEM}: None (0 bytes). Do not tile data into CMEM.
              \item \texttt{Vmem} (Vector Memory): 128 MB. Read: 18.5 TB/s, Write: 6.16 TB/s.
              \item \texttt{HBM} (High Bandwidth Memory): $\sim$17.2 GB at 820 GB/s.
          \end{itemize}
          
          \item \textbf{Interconnect:} 2D Torus, 4 links/chip, 45 GB/s per link.
      \end{itemize}
      
\noindent $\bullet$ \textbf{Optimization Strategies Required:}
      
      \begin{itemize}
          \item \textbf{Compute Mapping:} Map matrix multiplications to the Systolic Array (MXU). Map element-wise, reduction, and non-matmul ops to the Vector Unit (VPU).
          \item \textbf{Memory Tiling \& Fusion:} All functional tiling must target the 128 MB Vmem. Perform loop fusion to avoid spilling back to HBM.
          \item \textbf{Precision Tuning:} Evaluate suitability for \texttt{int8} quantization for 2$\times$ throughput.
          \item \textbf{Pipeline Parallelism:} Overlap HBM-to-Vmem DMA transfers with VPU/MXU execution.
      \end{itemize}
      
\noindent $\bullet$\textbf{Output Requirements:} Generate optimized code/IR and summarize: theoretical peak utilization, memory bandwidth utilization, and a bottleneck analysis.




\end{document}